\documentstyle[mprocl]{article}

\begin{document}

\title{DILUTE BOSE-EINSTEIN CONDENSATE IN A TRAP:
CHARACTERISTIC LENGTHS AND CRITICAL VELOCITIES}

\author{ALEXANDER L. FETTER}

\address{Departments of Physics and Applied
Physics, Stanford University\\Stanford, CA
94305-4060, USA\\E-mail:
fetter@leland.stanford.edu}

\maketitle\abstracts{
The Bogoliubov approximation and the Gross-Pitaevskii
equation characterize the effect of repulsive interactions
on a dilute ideal Bose-Einstein  gas in a spherical harmonic
trap. For large $N$, the interactions expand the condensate
relative to an ideal Bose gas; both the speed of sound and
critical angular velocity
$\Omega_{c1}$ for  creation of a quantized vortex depend
crucially on the interparticle repulsion through the
coherence length $\xi$. }

\section{Ideal Bose gas}

An ideal Bose gas provides a valuable introduction to the
physics of a real dilute Bose gas,  for it emphasizes
the fundamental role of the coherent  condensate
that contains a macroscopic number of particles in a
single quantum state.
The standard  example treats $N$ noninteracting bosons
in a volume
$V$ with periodic boundary conditions  and
uniform density $n = N/V$.
  In the classical limit,
the thermal de~Broglie wavelength $\lambda_T \equiv
(2\pi \hbar^2/mk_BT)^{1/2}$ is much shorter than the
interparticle spacing
$l \sim n^{-1/3}$.   As the temperature falls, however, the
thermal de Broglie wavelength grows, and the system eventually
becomes degenerate when $\lambda_T\sim l$.  Equivalently,
Bose-Einstein condensation in a uniform ideal gas occurs
 at $T_c \sim \hbar^2n^{2/3}/k_Bm$.

The situation is somewhat different for an ideal Bose gas in a
harmonic trap (taken as isotropic  for simplicity) with
$V_{\rm trap}(r) = {1\over 2} m\omega_0^2r^2$;  the
corresponding  oscillator length
$d_0~=(\hbar/m\omega_0)^{1/2}$ characterizes the size of the
(Gaussian) ground state.  In the classical limit ($k_BT \gg
\hbar\omega_0$), the density follows  the Boltzmann
distribution
$n_{cl}(r)~\propto~\exp[-V_{\rm trap}(r)/k_BT]$, which
can be rewritten as
$\exp(-r^2/2R_T^2)$, with $R_T \equiv
d_0(k_BT/\hbar\omega_0)^{1/2}$ the classical thermal
radius of the trapped gas (note that $R_T \gg d_0$).

 To
estimate the temperature $T_c$ for the onset of
Bose-Einstein condensation in a trap, use the same
  expression with
$n\sim N/R_T^3$, so that $k_BT_c \sim \hbar^2
N^{2/3}/mR_T^2\sim
\hbar^2\omega_0^2N^{2/3}/k_BT_c$; equivalently,
$k_BT_c\sim N^{1/3}\hbar\omega_0\gg \hbar\omega_0$.
Since $d_0\ll R_T$, the
condensate forms a narrow spike of width $d_0$ superimposed
on the smooth background of width $R_T$, as is seen clearly in
the original experiment   with a few
thousand $^{87}$Rb atoms.\cite{And}  More recent
experiments have condensed  millions of atoms, with typical
 parameters
$N\approx 5\times 10^6$,
$\omega_0/2\pi
\approx 120$ Hz, and $ d_0\approx 1.9~\mu{\rm m}$ for
sodium atoms.\cite{Dav}

\section{Effect of Repulsive Interactions}

The interparticle spacing is typically large compared to
the range of the atomic interactions, so that the
interparticle potential can be taken as $U(r)\approx
U_0\delta ({\bf r})$, with $U_0 = 4\pi a \hbar^2/m$ and
$a$ the $s$-wave scattering
length ($a~\approx~2.75$ nm for sodium atoms). The
Bogoliubov approximation$\,$\cite{Bog} assumes that nearly all
particles are in the condensate;
at zero temperature, the system  forms a dilute Bose gas
with
$na^3\ll 1$.

\subsection{Quasiparticles and the Coherence Length}

A uniform Bose gas at $T= 0$ K has only two
characteristic single-particle energies:  the
kinetic energy $\epsilon_k = \hbar^2k^2/2m$ and
the ``Hartree'' energy
\begin{equation}V_H({\bf r}) = {\textstyle\int} d^3r'\,U({\bf
r - r}') n({\bf r}') = U_0n = 4\pi
a\hbar^2n/m.\end{equation}
 Bogoliubov showed that a quasiparticle with wave vector $\bf
k$ has an energy$\,$\cite{Bog}
\begin{equation}E_k = \sqrt{2V_H\epsilon_k
+ \epsilon_k^2} \approx \cases{ \hbar s k,&for
$k\to 0$ (phonons),\cr
\epsilon_k=\hbar^2k^2/2m,&for $k\to \infty$ (free
particles),\cr}
\label{eq:NNB}\end{equation}
where $s^2 = U_0n/m = 4\pi a \hbar^2n/m^2$ is the
squared speed of sound. The quasiparticle is a
linear combination of a particle and a hole.  For $k\to 0$, it
is an equal particle-hole admixture, but it becomes a pure
particle as $k\to \infty$, with the cross-over  occurring at
$k\approx \xi^{-1}$, where $\xi =(8\pi a n)^{-1/2}$ defines
the ``coherence'' length.  Note that the speed of sound can
be rewritten as $s =
\kappa/2\pi\sqrt 2 \xi$, where $\kappa = h/m$ is the
quantum of circulation.  When $na^3\ll 1$, it is easy to
verify that $\xi
\gg l\gg a$.

\subsection{Radius of Interacting Condensate in a Trap}

The presence of the trap introduces a third
characteristic energy, and the Gross-Pitaevskii (GP)
equation$\,$\cite{EPG,LPP} for the nonuniform condensate wave
function $\Psi({\bf r})$ has the form
\begin{equation}(T + V_{\rm trap} +
V_H)\Psi = \mu\Psi,\label{eq:GP}\end{equation}
where $T=-\hbar^2\nabla^2/2m$ is the kinetic energy
operator, $V_{\rm trap}({\bf r})  = {1\over
2}\hbar\omega_0(r/d_0)^2$ is the trap potential energy,
$V_H({\bf r}) = U_0n({\bf r}) = 4\pi a \hbar^2n({\bf
r})/m$ is the Hartree potential energy of one particle with
the remaining particles, and
$\mu$ is the chemical potential.

For very low density, the coherence
length $\xi$ exceeds the trap size $d_0$, and the system acts
like an ideal Bose gas.
As
$N$ increases, however,
$\xi$ shrinks;  when $\xi$ becomes comparable with
$d_0$, the repulsive interactions
begin to  predominate, and the self-consistent radius
$R_0$ of the condensate  expands beyond  $d_0$.

To estimate the actual radius, note that the kinetic energy
per particle is of order
$T\sim\hbar^2/mR_0^2\sim\hbar\omega_0/R^2$, where $R
\equiv R_0/d_0$ is the dimensionless radius of the
condensate.  Similarly, the order of magnitude of the
remaining single-particle energies are
$V_{\rm trap}\sim \hbar\omega_0 R^2$, and
\begin{equation}V_H\sim
U_0n\sim{\hbar^2a\over m}n\sim
{\hbar^2a\over m}\,{N\over R_0^3}\sim
{\hbar\omega_0\over R^3}\,{Na\over d_0},\end{equation}
where  the dimensionless parameter $\eta \equiv Na/d_0$
characterizes the strength of the interactions.  For
$\eta\ll 1$, the system is effectively ideal, but for
$\eta\gg 1$, the interactions become crucial  (nevertheless,
the system remains dilute with $a\ll l$ for all practical
  trapped condensates).  If $\langle\cdots\rangle\equiv\int
dV\Psi^*\cdots\Psi$ denotes a condensate ground-state
expectation value, the total energy
$E=\langle T+ V_{\rm trap}+ {1\over 2}V_H\rangle$   is of
order
$\sim
N\hbar\omega_0(R^{-2} + R^2 + \eta\,R^{-3})$,
and $R$ is determined by minimizing $E$.

For small $\eta$, the minimum energy occurs at $R\approx
1$, and the condensate radius is just $d_0$, with $\mu
\approx {3\over 2}\hbar\omega_0$ (the ground-state energy
of the isotropic oscillator). For large
$\eta$, in contrast,  the kinetic energy is negligible, and
the minimum total energy occurs for $R^5\sim \eta$.
In this limit, a detailed calculation$\,$\cite{BP} shows that
$\mu = {1\over 2}\hbar\omega_0\,R^2$ and that
$R\equiv R_0/d_0 = (15\eta)^{1/5}$, which quantifies the
 expansion  of the condensate relative to $d_0$.  For
parameters appropriate to the MIT experiments,\cite{Dav} the
dimensionless condensate radius is
$R\approx 10.2$, so that
$R_0 \approx 19.3~\mu\rm m$.

The resulting  ``Thomas-Fermi'' (TF)
approximation neglects the kinetic energy entirely, and
the GP Eq.~(\ref{eq:GP}) then
determines the condensate density by the condition
$V_{\rm trap}({\bf r}) + V_H({\bf r}) = \mu$,
showing that the resulting density profile is parabolic.
The corresponding central density  $n(0) =
R_0^2/8\pi a d_0^4= (15/8\pi)\,N/R_0^3$ serves to
define both the speed of sound
$s^2 = 4\pi a \hbar^2n(0)/m^2$  and the coherence
length
$\xi^2=1/8\pi a n(0)$ for a trapped condensate;  as a
corollary, the relation $\xi R_0 =d_0^2$  implies   that
$d_0$ is the geometric mean of $\xi$ and $R_0$.  The previous
expression
$s=\kappa/2\pi\sqrt2 \xi$ then shows that
 the speed of sound in a large trapped
condensate increases linearly with $R_0$, reflecting the
increased density.  For the MIT experiments,\cite{Dav} these
relations  yield
$n(0)
\approx 4\times 10^{20}~\rm m^{-3}$, $s\approx 1.04~\rm
cm/s$, and
$\xi
\sim 0.187~\mu\rm m$.

In the Bogoliubov approximation, most of the
particles remain in the condensate, which
requires$\,$\cite{Bog}
$[n(0)a^3]^{1/2}\ll 1$.  For a trapped condensate at $T = 0$
K, this condition of small total noncondensate number holds
for
$R_0 a/d_0^2\ll 1$.  When this dimensionless ratio becomes
comparable with 1, however, the
repulsive interactions are so strong that the total
noncondensate number becomes of order $N$
(this situation occurs in liquid helium, even at $T=0$ K).
Comparison with the TF expression
$\eta=Na/d_0
\sim R^5$ shows that the Bogoliubov approximation fails
for a trapped condensate when
$N\sim (d_0/a)^6$, which is far larger than any current
experimental value (for the MIT trap with sodium atoms,
this limit merely requires that $N\ll 10^{15}$).

\section{Critical Linear and Angular Velocities}

Landau's original explanation of superfluidity involved the
critical velocity $v_c$; it is the speed at which a moving
macroscopic impurity can create quasiparticles and thus lose
energy.  A simple analysis shows that $v_c$ is the minimum
value of the ratio $\omega_k/k$, where $\omega_k$ is the
dispersion relation for the quasiparticles. From this
perspective, a uniform ideal Bose gas has zero critical
velocity, for the dispersion relation is simply $\hbar
k^2/2m$.  In contrast, Eq.~(\ref{eq:NNB}) shows that a
uniform dilute  interacting Bose gas indeed has a nonzero
critical velocity, with $v_c = s\propto a^{1/2}$, arising from
the presence of the repulsive interactions.
\subsection{Critical Angular Velocities in a Trapped
Condensate}
 For a dilute trapped condensate, it is natural to
take the speed of sound $s$ as the critical velocity.
It is impractical to shoot an impurity through the
condensate, but the equatorial speed of a rotating
condensate can serve to define a corresponding critical
angular velocity
$\Omega_c = s/ R_0 =\hbar/md_0^2\sqrt2=
\omega_0/\sqrt2$. Thus a
large vortex-free   condensate rotating at an
angular velocity
$\Omega\le
\Omega_c$ should remain superfluid   (this expression also
indicates that the frequencies of   low-lying compressional
modes in a trapped condensate are independent of the radius
and of order
$\omega_0$).

The
low-lying hydrodynamic modes of a large spherical condensate
have the dispersion relation
$\omega_{nl} = \omega_0 [l + n(2n+2l+3)]^{1/2}$, where $n$ is
 the radial quantum number  and $l$ is the
orbital-angular-momentum quantum number.\cite{Str}  For fixed
$l$, the radial wavenumber $k$ is $\approx n/R_0$, and a
corresponding more precise Landau critical angular velocity
is
$\sqrt2 \omega_0$ (this value occurs as $k\to \infty$).

 It is helpful to review the
effect of rotation
${\bf
\Omega} = \Omega\hat z$ on a sample of liquid helium.  If the
fluid is normal, the microscopically rough walls  bring it
into solid-body rotation
${\bf v}_{\rm sb} = \bf \Omega\times r$, where $\bf r$ is the
distance from the axis of rotation; this flow has uniform
vorticity, with
$\nabla\times {\bf v}_{\rm sb} = 2\bf \Omega$.  In the
rotating frame, the walls are stationary, and the relevant
zero-temperature thermodynamic function is the ``free
energy'' $F = E-{\bf
\Omega \cdot L}$ where
$E$ is the total ground-state energy and $\bf L$ is the total
ground-state angular momentum. For a circular cylinder of
radius
$R_0$, the free energy of a state with one vortex on the
symmetry axis becomes lower than that with no vortex at a
lower critical angular  velocity
$\Omega_{c1}= (\kappa/ 2\pi
R_0^2)\,\ln(R_0/
\xi)$, where $\xi$
represents the vortex core radius ($\xi$ is a few atomic
diameters for superfluid ${}^4$He).

Assuming that a similar
expression holds for the creation of a vortex in a
large trapped condensate,\cite{BP,DS} the lower critical
angular velocity is smaller than the trap frequency $\omega_0
= \kappa/2\pi d_0^2$ by a factor of order
$(\xi/R_0)\,\ln(R_0/\xi)\ll 1$.  In addition to the
  compressional modes with frequencies of order
$\omega_0$, the presence of a vortex line
 introduces
new dynamical degrees of freedom associated with ``vortex
waves.'' At long wavelengths ($k\xi\ll 1$), the classical
vortex-wave dispersion relation
$\omega_k \approx (\kappa k^2/4\pi)\,\ln(1/k\xi)$ immediately
suggests  low-lying normal modes with $k\sim R_0^{-1}$
and frequencies of order $\Omega_{c1}$, which might serve to
signal their presence.

\subsection{Analogy with Type-II Superconductors}

These results for $\Omega_c$ and $\Omega_{c1}$
 are very similar to the thermodynamic  critical field $H_c =
\Phi_0/2\pi\sqrt2\lambda\xi$  and lower critical field
$H_{c1} \approx (\Phi_0/2\pi \lambda^2)\,\ln(\lambda/\xi)$,
where
$\Phi_0 = h/2e$ is the flux quantum, $\lambda$ is the
penetration length, and
$\xi$ is vortex core radius (also the superconducting
coherence length).\cite{Tink} A uniform bulk superconductor
becomes unstable with respect to normal metal at the field
$H_c$, and  the formation of quantized flux lines
(vortices) becomes favorable at the field
$H_{c1}$.  In a type-II superconductor (one with $\lambda
>\xi/\sqrt 2$), the thermodynamic
instability at
$H_c$ is preempted by vortex formation at $H_{c1}<H_c$.

The electromagnetic currents in a
charged superfluid cut off the logarithmic  intervortex
interaction
 potential, screening it   exponentially  beyond the
penetration length
$\lambda$.  This quantity diverges as the charge on each
particle tends to zero, and the corresponding
``screening'' length in a neutral superfluid becomes either
the radius of the container or the intervortex separation,
whichever is smaller.

 In addition, a type-II superconductor
 ultimately becomes normal at the upper critical field
$H_{c2} = \Phi_0/2\pi\xi^2$, roughly when the vortex cores
overlap.  Unfortunately, the corresponding $\Omega_{c2} =
\kappa/2\pi\xi^2$ is unattainably large in superfluid
${}^4$He, and mechanical instability
for $\Omega\ge \omega_0$ may  also render it  unobservable
in a rotating dilute trapped condensate (in this case,
a trapped condensate in equilibrium could contain only
 relatively few vortices).

\subsection{Non-Axisymmetric Rotating Traps}

The magnetic field that confines a  dilute atomic
Bose condensate acts simply as a potential $V_{\rm trap}({\bf
r})$; this situation
  differs greatly from the microscopically rough
walls of a container for superfluid helium.  Thus,  a
``rotating trap'' is meaningful only to the extent that it is
nonaxisymmetric, for the  rotating time-dependent potential
 pushes the condensate, setting it into  motion.

In contrast to the case of an axisymmetric potential such as a
circular cylinder,
 the free energy
$E-\Omega L$ of an irrotational vortex-free state in a
nonaxisymmetric trap decreases with increasing
$\Omega$ (like
$-{1\over 2}I\Omega^2$, where $I$ is an effective moment of
inertia).  In the limit of large distortion, $I$ can
approach that of solid-body rotation.  The negative  free
energy of the vortex-free state typically delays the onset of
vortex formation, for the
 vorticity localized in the vortex cores becomes less
necessary to
 ``mimic'' the uniform vorticity of ${\bf v}_{\rm sb}$.
This effect is readily verified in  simple cases.  For a
uniform superfluid in a long rotating elliptic cylinder with
semiaxes $a$ and
$b$,\cite{ALF}
 the preceding expression
$\Omega_{c1}=(\kappa/2\pi b^2)\ln(b/\xi)$  applies if $b=a$,
but
the corresponding lower critical angular velocity $\Omega_{c1}
\approx (\kappa/4\pi b^2)\ln(b/\xi)$ for $b\ll a$
 can become  significantly
larger.

The GP Eq.~(\ref{eq:GP}) provides a basis for analyzing
 a rotating trap, where the confining potential
$V_{\rm trap}({\bf r})$ is stationary in the rotating frame
 ($\bf r$ now denotes the coordinate in the rotating frame).
The free energy is given by $E - \Omega L_z$, where $E$ is
the ground-state energy (assuming $T = 0$ K for
simplicity).
The condensate wave function  $\Psi = |\Psi|e^{iS}$
can be expressed as a magnitude
$|\Psi|= n^{1/2}$ and a phase $S$ that determines the
superfluid velocity ${\bf v} = (\hbar/m)\nabla S$.  The TF
limit of a large condensate neglects the spatial variation of
$n$, and the free energy  becomes
\begin{equation}F\approx {\textstyle \int}
d^3r\,\big({\textstyle{1\over 2}}mnv^2 + nV_{\rm trap}
+{\textstyle{1\over
2}}U_0n^2-mn\Omega\hat z\cdot{\bf r\times
v}\big).\label{eq:GP2}
\end{equation}

In a singly connected container, $\bf v$ can be written as a
sum of two contributions:  ${\bf v}_\Omega\propto \Omega$
arising from the moving walls and ${\bf v}_\kappa$ arising
from the vortices with circulation~$\kappa$.
Correspondingly, $F$ separates into a
(vortex-free) contribution
$F_\Omega$ associated solely with ${\bf v}_\Omega$ and
$F_{\kappa} = \int d^3 r\,mn{\bf v}_\Omega\cdot{\bf v}_\kappa
+ {1\over 2}mnv_\kappa^2 - mn\Omega\hat z\cdot{\bf r\times
v}_\kappa$ that  also depends on the angular
velocity $\Omega$, both explicitly and through ${\bf
v}_\Omega$.  The critical angular velocity
$\Omega_{c1}$ for vortex creation occurs when
$F_{\kappa}$ first vanishes.

 The  integral for $F$ provides a  variational
expression in the rotating frame;  as an approximate trial
function for the phase, take
$S= -(\hbar/m) \Phi$, where $\Phi$ is the classical
velocity potential
for a uniform irrotational fluid (conventionally defined so
that
${\bf v}=-\nabla\Phi$).  This velocity field is
automatically  irrotational apart from the vortex cores and
satisfies the condition that its normal derivative match
the normal velocity of the rotating boundary. In addition,
$\nabla\cdot{\bf v}= 0$,  which is the appropriate TF limit
of the continuity equation
$\nabla\cdot(n{\bf v}) = 0$.  The actual trapped-condensate
density differs from that for $\Omega = 0$ only because of
the Coriolis and  centrifugal forces of order
$m{\bf\Omega\times v}$ and $m{\bf \Omega}\times
({\bf\Omega\times  r})$, respectively.  Each of these is
small compared to the force of the harmonic trap so long as
$\Omega\ll\omega_0$, in which case it is  permissible to
neglect the change in density (apart from the formation of
the small vortex core of radius $\approx
\xi$).

As a result, the  free energy $F_\Omega$ for an  irrotational
condensate in a rotating trap    has the same form as that for
a uniform incompressible irrotational fluid, but with the
particle density taken from  the  solution of the GP
equation.  Since the actual density varies slowly in the
TF limit, $F_\Omega$ differs from the classical expression
only through factors of order unity, representing various
moments of the actual density.  Similarly,    the additional
free energy     $F_\kappa$ arising from the presence of a
vortex line  also  differs from the classical
expression for a uniform incompressible fluid only by factors
of order unity, because the TF density profile cuts off the
logarithmic divergence in the kinetic energy at a distance of
order
$\xi$, which can be identified with the classical vortex
radius. Consequently, the classical analysis$\,$\cite{ALF}
for uniform fluid in a rotating elliptical cylinder provides
a  qualitative guide to the corresponding problem of vortex
nucleation in a rotating elliptical trap.

This correspondence remains valid for a multiply connected
container, when the velocity field $\bf v$   includes an
additional  contribution ${\bf v}_\Gamma$ arising from the
quantized circulation $\Gamma = j\kappa$ around the various
internal boundaries (here, $j$ is an integer). For example,
consider an incompressible fluid in an asymmetric annular
region between two nonconcentric cylinders (outer radius
$R_2$ and inner radius $R_1<R_2$), with the centers displaced
a distance
$d\le R_2-R_1$. If the system rotates in equilibrium at
angular velocity $\Omega$ around the symmetry axis of the
inner cylinder,   the critical angular velocity
$\Omega_\kappa$  for  the creation of vortex-free  quantized
circulation
$\kappa$ around the inner boundary can be evaluated
for arbitrary $d$.  In the symmetrical limit ($d\to 0$), the
result
$\Omega_\kappa = (\kappa/2\pi)
(R_2^2-R_1^2)^{-1}\ln(R_2/R_1)$ is similar to that for a
cylinder;  in the small-gap limit, however,  the resulting
expression
$\Omega_\kappa
\approx
\kappa/4\pi R_1R_2$ has no logarithmic factor.  These
expressions  can serve to estimate the corresponding
quantities for a trapped toroidal condensate (created, for
example,  by piercing the condensate with an off-axis laser
beam).\cite{MRA} The circulation-induced deformation of
the condensate might serve to detect the presence of a
persistent current.

\section*{Acknowledgments}
I am grateful to M.~R.~Andrews and D.~Rokhsar for
valuable discussions and insights.  This research is supported
in part by the National Science Foundation, under Grant
No.~DMR 94-21888.

\eject

\end{document}